# *The full detector simulation for the ATLAS experiment: status and outlook*

A. Rimoldi
*University of Pavia & INFN, Italy*
A.Dell'Acqua
*CERN, Geneva, CH*

The simulation of the ATLAS detector is a major challenge, given the complexity of the detector and the demanding environment of the LHC. The apparatus, one of the biggest and most complex ever designed, requires a detailed, flexible and, if possible, fast simulation which is needed already today to deal with questions related to design optimization, to issues raised by staging scenarios, and of course to enable detailed physics studies to lay the basis for the first physics discoveries. Scalability and robustness stand out as the most critical issues that are to be faced in the implementation of such a simulation. In this paper we present the status of the present simulation and the adopted solutions in terms of speed optimization, centralization of services, framework facilities and persistency solutions. Emphasis is put on the global performance when the different detector components are collected together in a full and detailed simulation. The reference tool adopted is Geant4.

## 1. INTRODUCTION

The simulation of the Atlas experiment is an ambitious task since it requires a careful and detailed design for providing optimal functionalities in the different domains characterizing it.

Its global design is indeed very complex since the detector to be simulated is out of scale respect to any previous ever built: the local environment is demanding because the biggest collaboration of physicists ever gathered is faced to the most complete and challenging physics scenario of the LHC. The simulation domain in Atlas wraps as its own components:

- the fast simulation project
- the digitization for the implementation of the different detector responses
- the MonteCarlo generators
- the simulation in GEANT3, operatonal since ten year and still used for Data Challenges purposes
- the new simulation in GEANT4, under imple-mentation and test and ready to be used as Atlas baseline starting from mid 2003.

In particular the new simulation in GEANT4 has been developed and tested for physics validation studies to verify the physics content in GEANT4, it was then carefully compared with the current simulation performed in GEANT3 and with real data from testbeam experimental setups, where modules from the subdetectors productions are tested on beam lines.

The new simulation should also account for staged scenarios and to beam or machine constraints at the LHC startup.

Due to the huge amount of details characterizing it and to the need of improving performances, the simulation itself should also house a fast-simulation for parts of the detector or should gain from the implementation and use of parameterization techniques for optimization issues, without loosing the benefits of a detailed geometry description.

In this paper is presented an overview of the simulation project in Atlas with emphasis to the most recent applications in the GEANT4 geometry description of the whole detector and the testbeam setup implementations, while physics validation issues are presented in a different CHEP03 paper (Ref.1).

## 2. A ROAD TO THE ATLAS SIMULATION

Data Challenges (DC) are the software milestones for the experiment and simulation is playing the main role for generating the requested event samples. Since the end of 2001 many tests on event productions were performed with the standard GEANT3 simulation. This phase, called DC0, was meant for resuming all the functionalities in place at the Physics Technical Design Report (TDR) time (1999) and to start the procedure for further DCs in the same environment with improved functionalities or using new tools.

During DC0 the GEANT4 Atlas community performed a dedicated production of single particle and Higgs events (~10**6) for a selected geometry configuration as early test for application robustness and distributed site production (CERN and Japan).

The apparatus geometry described in this application contained a detailed description of the muon system and the tile calorimeter, while rough description of other parts of the Atlas detector were fit together.

The following phase (DC1), done in the GEANT3 environment, was characterized, at the beginning, by event generation of small event samples (single particle, B scans and Higgs), then it continued through high statistics single particle, minimum-bias events, QCD-jets and physics events productions.

These samples were requested by the different Atlas physics communities for testing purposes in view of the preparation of the subdetectors Technical Design Reports.

During this phase, lasting 2002 and continuing in 2003, checks of robustness for large production samples were performed.

For DC2, expected at startup in late fall 2003, a heavy plan of event productions will undergo both in GEANT3 and in GEANT4 frameworks with goal to refine the present procedures and to add functionalities, being the basis for the Computing TDR (2004).

Since May 2000 an extensive programme of physics validations in GEANT4 was launched in order to authenticate the physics content of GEANT4.

The purpose was twofold: for the GEANT4 physics benchmarking and the GEANT4 physics validation itself. The aim was to compare features of interaction models with similar features in the Geant3.21 baseline and try to understand differences in applied models, e.g.: effect of cuts on simulation parameters (range cut vs. energy threshold, for example).

For the physics validation itself, the purpose was related to the use of available experimental references from testbeam for various sub-detectors and particle types to determine prediction power of models, to estimate the Geant4 performance using different sensitivities of sub-detectors (energy loss, track multiplicities, shower shapes…), to tune Geant4 models (“physics lists”) and parameters (“range cuts”) for optimal representation of the experimental detector signal.

The GEANT4 simulation and its validation using testbeam results is the only reality check until there is real data to compare with.

## 3. SIMULATION DATA FLOW

In the ATLAS experiment the simulation data flow is designed in order to optimize the link between the different components as shown in Fig.1.

In the simulation there are a number of steps that need to be undertaken. The passage of particles through the detectors is recorded as a number of positions throughout the detectors themselves.

The hits information is combined with estimates of internal noise and subjected to a parameterization of the known response of the detectors to produce simulated digital output (digits). The digits can be fed to the pattern recognition and track reconstruction algorithms as if they were real data.

The diagram presents the simulation data flow, including Generator and HepMC, pile-up and digitization when the hits generation is performed by the simulation itself and merged hits, after pile-up or from the Level1 digitization, are passed to the ROD emulation algorithm for further processing.

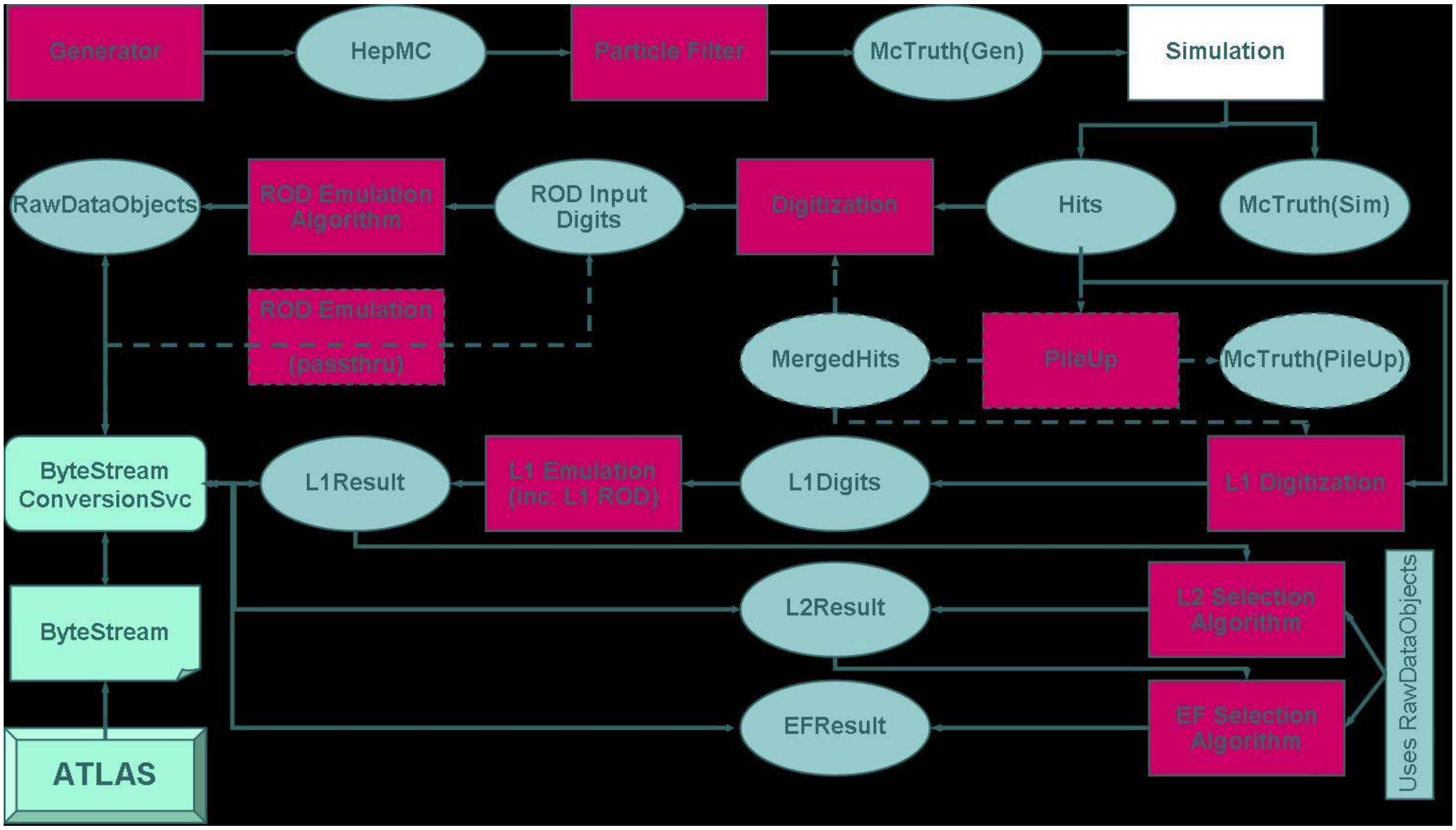

Fig.1 The Atlas Simulation data flow

## 4. THE GEANT3 SIMULATION

The GEANT3 simulation is the baseline for Atlas since more than ten years; the package was frozen since 1995 and it now is representing the current simulation for the experiment.

The program parameters characterizing it are an operative expression of the expected complexity for this experiment:

- 27 Millions of distinct volumes copies
- 23 thousand of different volume objects

The processing time per job is about 24 hours while the typical output file for a production of a sample of ~170-320 events (comprehensive of hits and digits) is ~200-300 MB.

The detector components are simulated in an evolving representation, while they have been in a stable configuration during all DC1 productions.

Pile up was used in the latest production, but full pile-up with cavern background should be added.

The simulation program using GEANT3 is still in evolution. In its latest version new implementations were made in the pixel detector and in the forward moderator. Inactive material description and forward shielding are simulated and the correspondent material is taken into account for background evaluation.

## 5. THE GEANT4 SIMULATION

Since years a big effort was devoted to the developing of the simulation program in the new OO environment. In all the Atlas subdetectors a wide program of implementation of a correct, complete and detailed simulation was launched and most applications are already implemented in the common framework of Atlas, among them the complete geometry description of the whole apparatus.

The new Geant4 based simulation is meant for dealing with the new physics environment, with extensive tests on apparatus prototypes against real data from testbeam before and during the massive productions of the whole detector components for the real experiment.

Currently most applications are using the GEANT4 version 4.1.1 and the upgrade to the new version (5.1R2) was successfully made for the geometry part in all the detector components in the standard framework.

### 5.1.Framework

The simulation makes use of the standard Atlas framework (Athena-Gaudi) for all the domain's components and also profits from standalone applications for development purposes. For the GEANT4 based simulation most applications are developed in a the specific framework, FADS/Goofy (Framework for Atlas Detector Simulation in a Geant4 OO FollY).

The framework, born three years ago under redhat6.2 and gcc compiler version 2.95, is successfully running since more than one year in Redhat 7.2, gcc 2.96; the migration to Redhat 7.3.1 gcc 3.2 was successfully completed in 2003.

The FADS/Goofy framework was widely used for subdetector applications in the development phase or for testbeam developments implying physics validation studies.

This framework is very light, portable and fast. The dynamic loading facility is extensively used as well as the lazy instantiation technique, that allows performing actions on demand through uploading or downloading of specific libraries.

In that way the possibility of collecting detector elements at run time, without a predefined detector structure is easily possible.

After the development in the simulation framework, the subdetector applications were implemented in the standard Atlas framework (Athena-Gaudi) without any modification.

### 5.2. Detector geometry in GEANT4

The four main subdetectors composing Atlas are implemented in the new simulation with a high level of details.

Figures 2 to 6 show three-dimensional views of some geometry implementation of the Atlas subdetectors: the SCT component of the Inner Detector is shown in Fig.2, the LAr calorimeter module in Fig.3, while the complete Tile calorimeter in Fig.4.

The muon system is shown in Fig.5: a cut view shows the four different components characterizing it: the precision chamber system in the barrel and forward region (MDT and CSC chamber system) and the trigger system (RPC and TGC system); the complete geometry for the barrel and endcap toroids in shown in Fig.6.

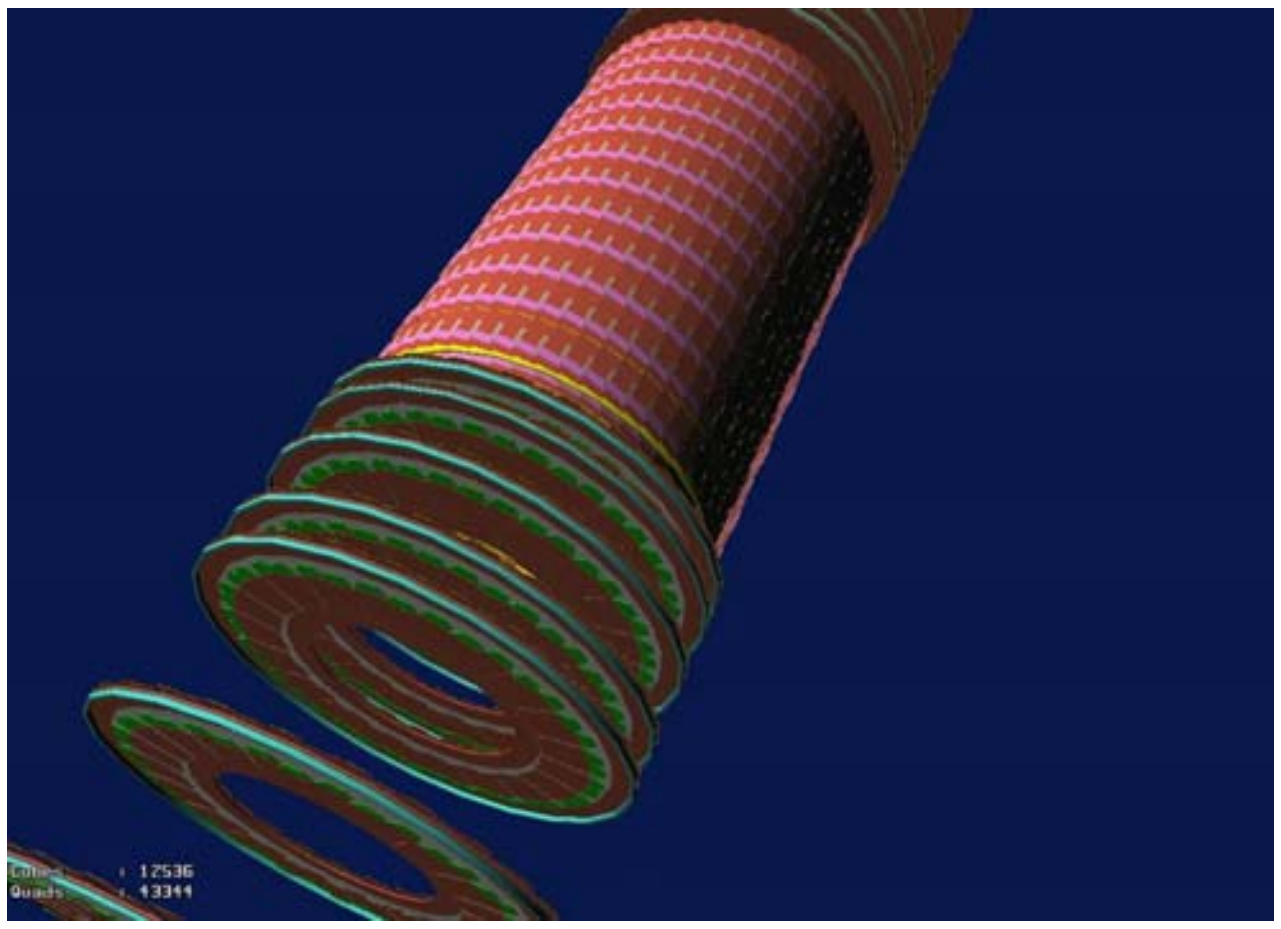

Fig. 2 The SCT detector of Atlas

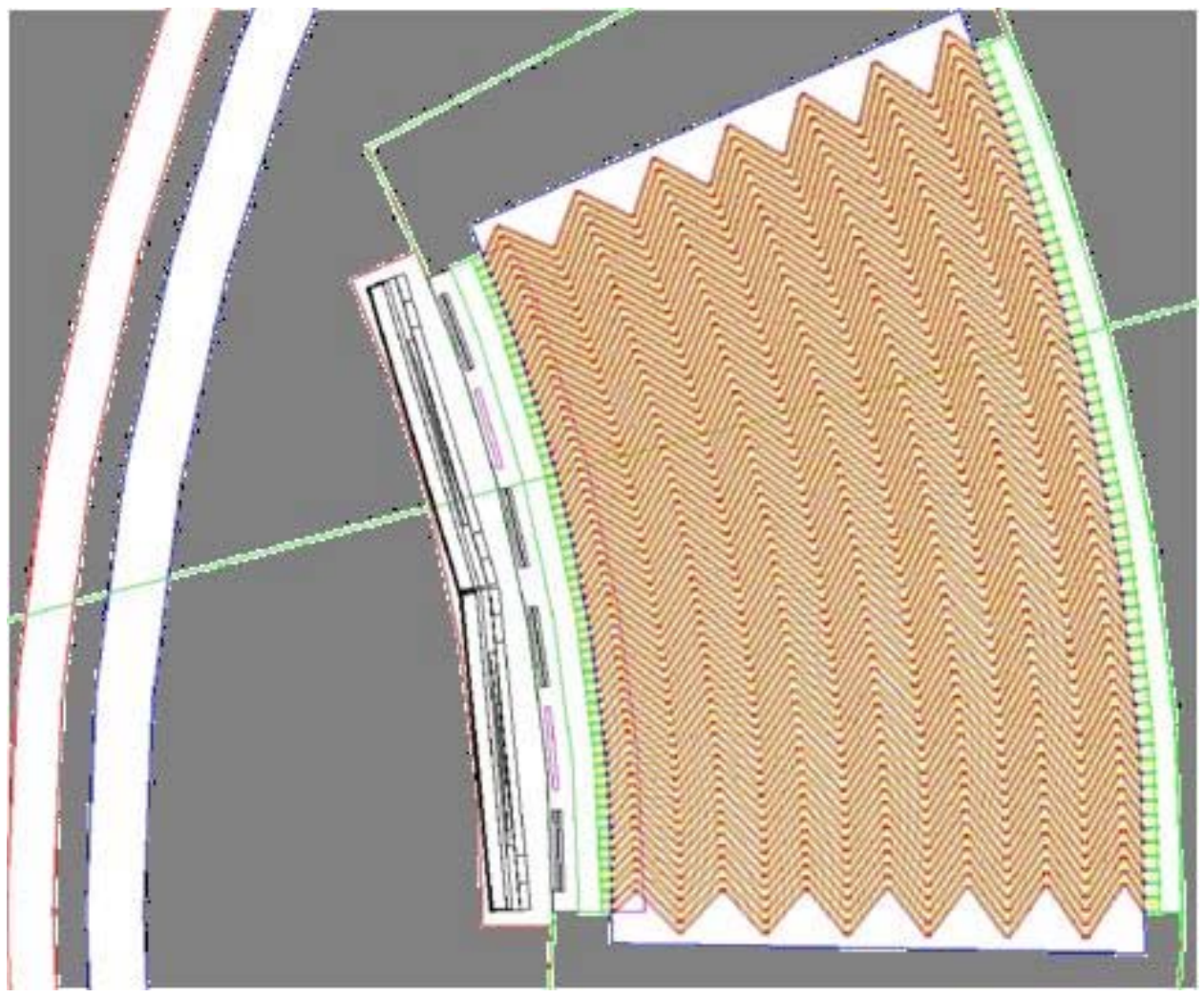

Fig.3 LAr electromagnetic calorimeter testbeam module

The level of details reached by the new simulation in the geometry description is at least of the same order of magnitude with respect to the one achieved in the previous simulation in Geant3, while the implementation details differ in some cases from the previous simulation due to new functionalities in GEANT4.

In the muon system, for example, special solid hexagonal shapes have been implemented in order to describe the CSC

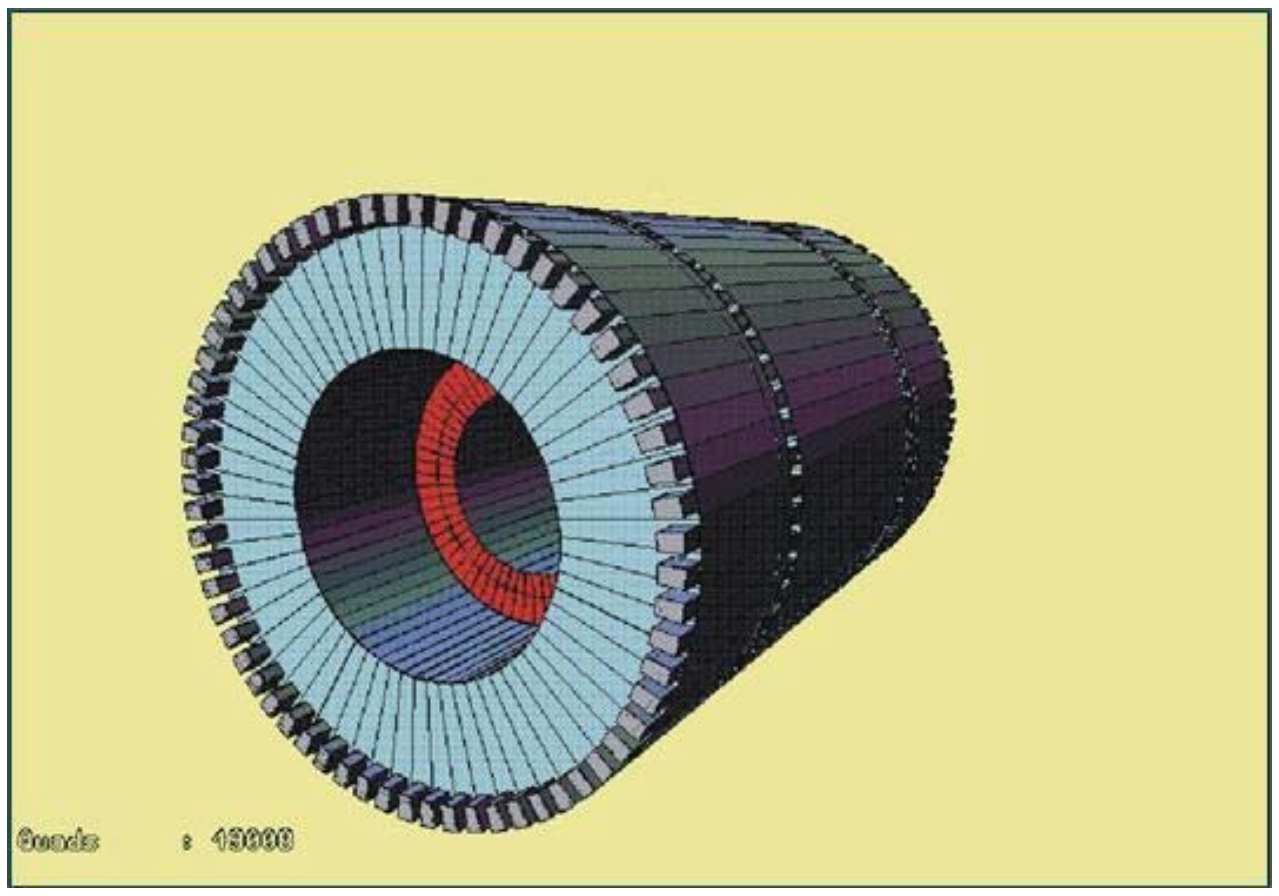

Fig.4 The tile calorimeter three dimensional view.

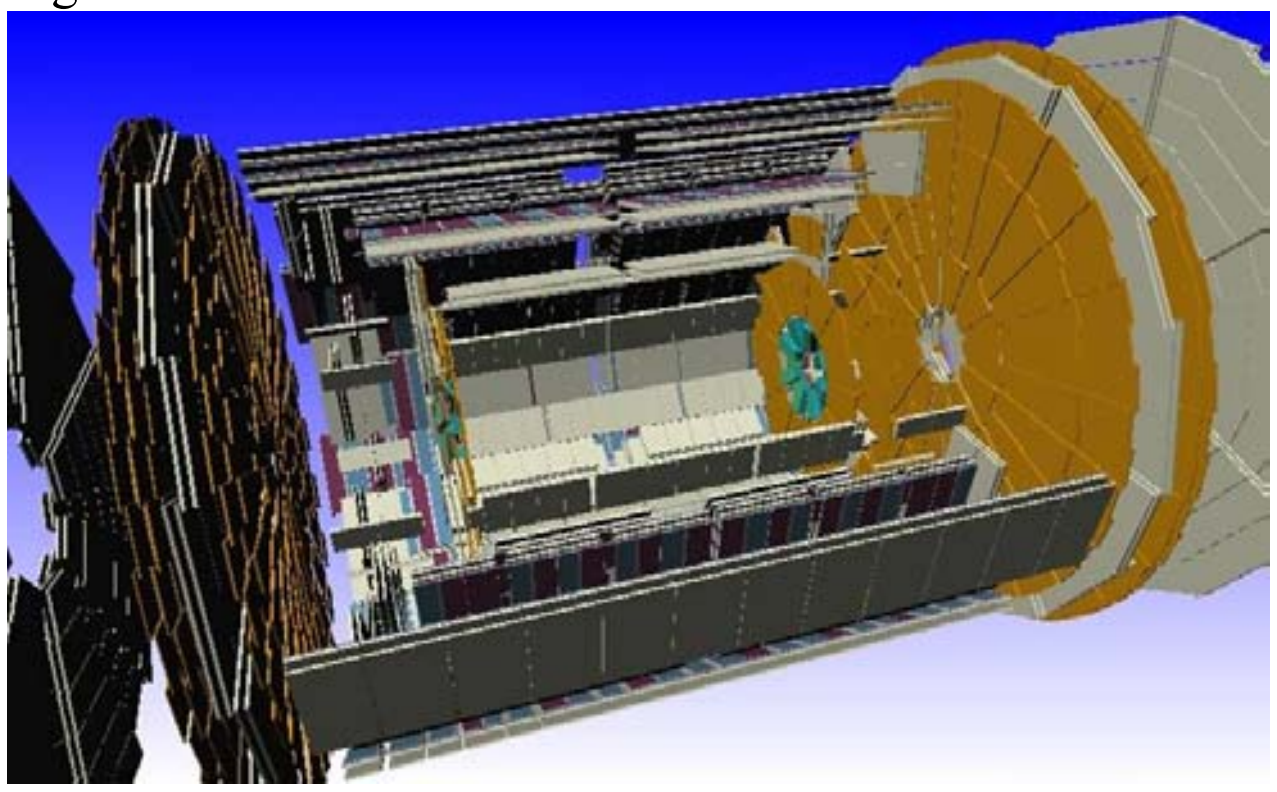

Fig.5 Open view of the muon system with all the precision system and the trigger system (barrel and endcap regions)

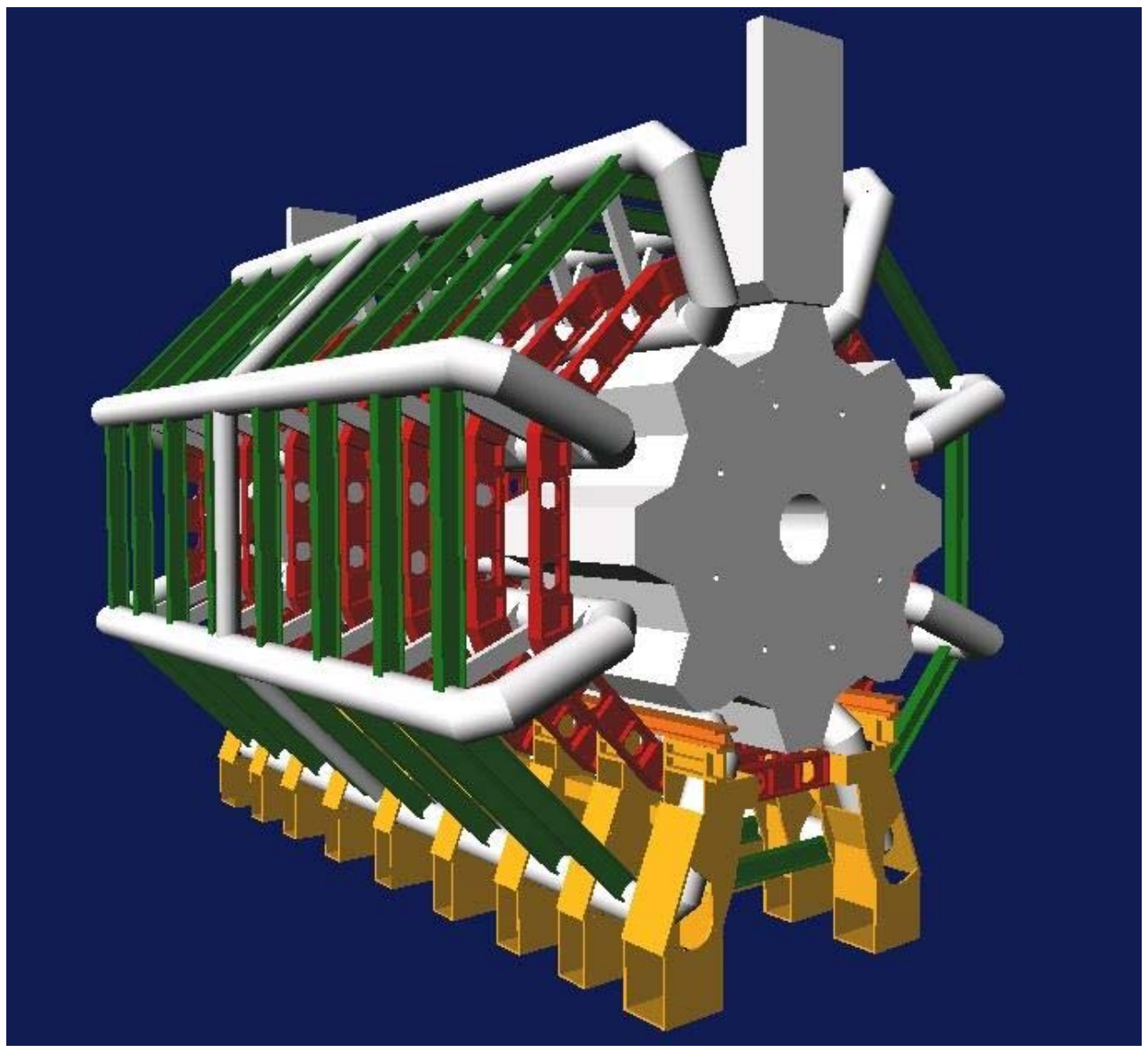

Fig.6 The Atlas toroid system with feet, rails and service chimneys.

system in the forward region, for the toroids implementation wide use of boolean solids was adopted and for the description of the LAr calorimeter new solid shapes describing the 'accordion' geometry were built.

## 5.3.Testbeam geometry in GEANT4

In the past years an extensive test of modules in each subsystem was performed in dedicated setups on beam lines and a parallel and wide effort in the simulation followed, in order to verify the physics content of GEANT4.

Comparisons with the previous simulation in GEANT3 were performed, for tuning the new simulation and understanding the possible differences when tuning was ineffective.

For each subdetector setup a detailed simulation in GEANT4 was put in place. Emphasis was put in the geometry description and in the implementation of the beam line characteristics (beam momentum tuning, beam spread, vertex displacement from the nominal position).

Fig. 7 shows the geometry of the tile testbeam modules as simulated after the 1996 data taking period (upper part of Fig. 7) and in a later setup (2001) (lower part of Fig.7).

Fig.8 presents a three dimensional view of the complex setup for the forward calorimeter along the beam line: cryostat containing the calorimeter modules and all the components on the setup are described in the simulation.

In Fig.9 the setup for the combined geometry of pixel detector, tile and muon system sector is shown.

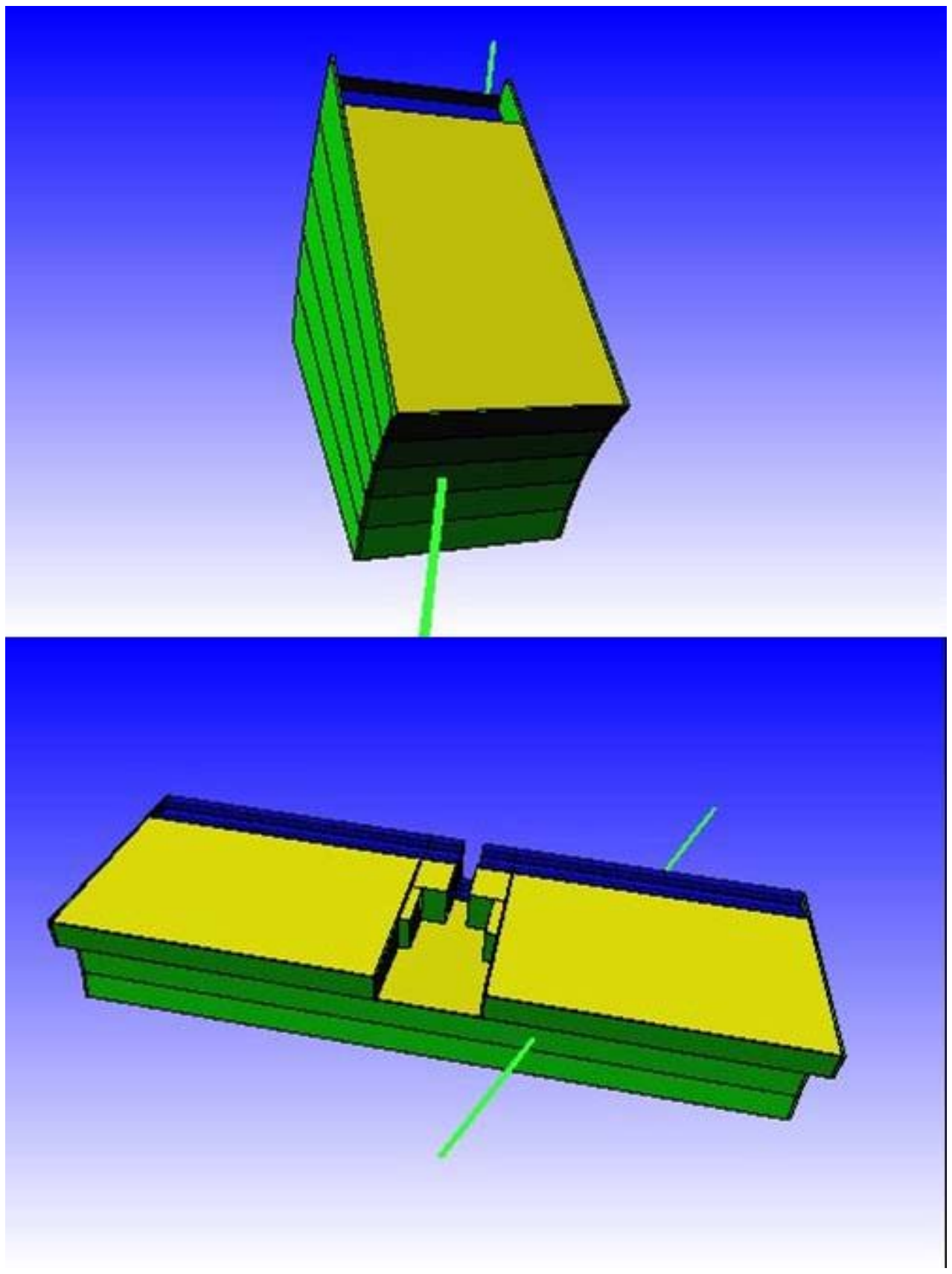

Fig.7 three dimensional view of the tile testbeam modules as in the 1996(up) and 2001(down) setups.

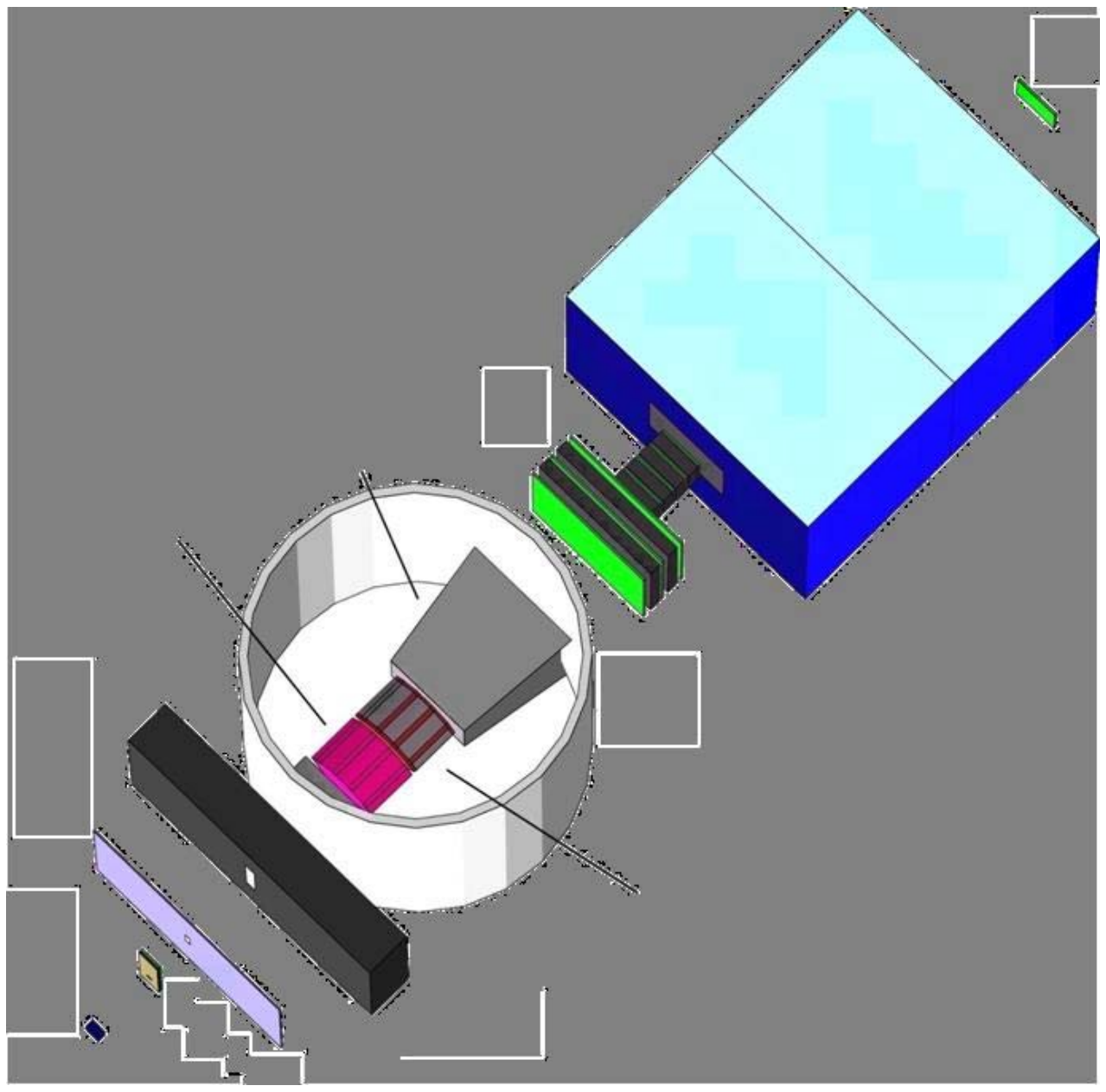

Fig.8 The forward LAr calorimeter (FCAL) testbeam setup

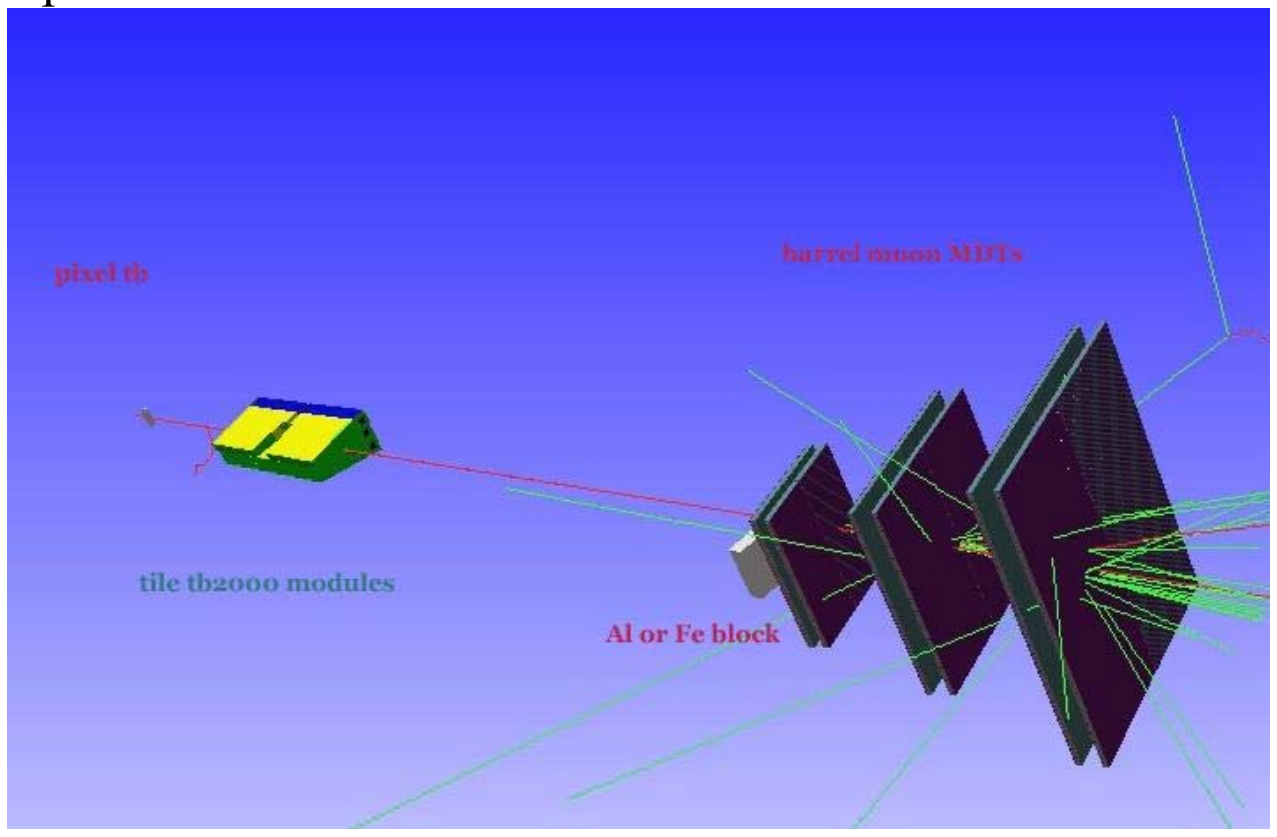

Fig.9 Testbeam setup for the combined configuration as in CERN H8 testbeam area (summer 2002) with Pixel detector elements, tile calorimeter and barrel sector of the muon system with some extra material in front of the muon setup (Al or Fe blocks).

Using these geometries we approached the physics validaton studies in the different Atlas subdetectors: we compared features of interaction models with similar features in Geant 3.21baseline. For the validation we used all available experimental references from testbeam and different sensitivities of subdetectors to estimate the Geant4 performance and we tuned physics lists and parameters (range cut) for optimal representation of the experimental detector signal.

## Acknowledgments

The authors wish to thank all the contributors, developers and users: they made possible the realization of the Atlas simulation since the early times of the Atlas life.

Their comments were essential for refinements to the final design and for developing new features that made possible the realization of this work. The following list of references documents the huge amount of work behind this short presentation (Ref. 2 to 8) performed in the last years from a community involving more than 20 people.